\documentclass[12pt]{article}
\usepackage{amsmath,amssymb,graphicx,mathrsfs,slashed}
\usepackage[usenames, dvipsnames]{color} 
\usepackage{tabularx}
\usepackage[colorlinks=true]{hyperref}
\hypersetup{
	colorlinks=true,
	linkcolor=black,
	filecolor=magenta,      
	urlcolor=cyan,
	pdftitle={Sharelatex Example},
	bookmarks=true,
	pdfpagemode=FullScreen,
	citecolor=black,
}
\usepackage{geometry}
\newcommand{\be}{\begin{equation}}
\newcommand{\ee}{\end{equation}}
\newcommand{\bea}{\begin{eqnarray}}
\newcommand{\eea}{\end{eqnarray}}

\def\({\left(} \def\){\right)}

\newcommand{\al}{\alpha}

\begin{document}
\title{\vspace{-0.9in}
{\textbf{Hyperbolicity Constraints in Extended Gravity Theories}}}

\author{\large Yotam Sherf
	\\
	\vspace{-.5in}  \vbox{
		\begin{center}
			$^{\textrm{\normalsize
					\ Department of Physics, Ben-Gurion University,
					Beer-Sheva 84105, Israel}}$
			\\ \small 
			 sherfyo@post.bgu.ac.il
		\end{center}
}}
\date{}
\maketitle
\begin{abstract}


  We study the characteristic structure of the Einstein-Hilbert (EH) action when modifications of the form of $R^2,~ R_{\mu\nu}^2$, $R_{\mu\nu\rho\sigma}^2$ and $C_{\mu\nu\rho\sigma}^2$ are included. We show that when these quadratic terms are significant, the initial value problem is generically ill-posed
  . We do so by demanding the hyperbolicity of the effective metric for propagation of perturbations. Here, we find a general expression for the effective metric in field space and  calculate it explicitly about the cosmological Friedman-Robertson-Walker (FRW) spacetime, and the spherically symmetric Schwarzschild solution. We find that when these quadratic contributions are non-negligible, the signature of the effective metric becomes non-Lorentzian and hence non-hyperbolic. As a consequence, we conclude that theories suggesting the inclusion of these terms can only be considered as a perturbative extension of the EH action and therefore cannot constitute a true alternative to general relativity (GR).

\end{abstract}

\renewcommand{\baselinestretch}{1}\normalsize
\newpage
\newgeometry{margin=1in,top=1in}

\section{Introduction}\label{2}


Large distance scales phenomena are well described in the frame of 
General Relativity. While at short distances, the classical GR description is incomplete, indicating the need in modifications of the EH action. Among them, the modifications that are mostly expected to appear, are higher derivative terms in the form of higher powers of the curvature tensors \cite{Clifton:2011jh,Salvio:2018crh,Salvio:2017xul}. 

Higher derivative theories are relevant in several contexts. First, in string theory where higher derivative terms naturally appear in its low energy limit \cite{Myers:1987yn,Callan:1985ia} and in the study of black hole (BH) solutions \cite{Lu:2015cqa,Stelle:1977ry}. Here we are mainly interested in their relevancy in the context of low curvature backgrounds such as the cosmological FRW spacetime \cite{Clifton:2011jh,Salvio:2017xul,Carroll:2004de,Nojiri:2010wj,Starobinsky:1980te,DeFelice:2010aj,Alvarez-Gaume:2015rwa,Nojiri:2006ri,Nojiri:2001ae}, where in these models, modifications to GR are studied in order to explain the dark energy,  early-time inflation with late-time cosmic acceleration, and cosmological perturbations. 

Here we would like to examine the possibility that higher derivative  theories can be viewed as an alternative to the Einstein theory. This implies that the additional terms induce significant modification to the classical GR also in large distance scales. Otherwise, they can be viewed as a perturbative correction to the GR \cite{Camanho:2014apa}. Then, we study the violation of causality by investigating the hyperbolicity of the modified EOMs of perturbations. We assume that the contribution of the higher derivative terms can be comparable or larger than that of the Einstein term. as we explain in detail below.


We consider modifications to the EH action in four dimensions (4D) of the form\footnote{The case of the quadratic Weyl tensor will be discussed separately.}
\begin{equation}
\mathcal{S}=\dfrac{1}{16\pi l_p^2}\int d^4x\sqrt{-g}\big({R}+\lambda_1{R}^2+\lambda_2{R}^{\mu\nu}{R}_{\mu\nu}+\lambda_3{R}^{\mu\nu\rho\sigma}{R_{\mu\nu\rho\sigma}}\big)~,
\label{1.1}	
\end{equation}
where the $\lambda_i$ are dimensionful coupling constants of a typical length scale $\lambda_i\sim l^{2}_i$. These coefficients determine the length scale at which the corrections are important. Obviously, when these coefficients are significantly small ${l_i^2}/{l_P^2}\ll1$ or at low curvature backgrounds, the contribution of the additional curvature terms is negligible and can be considered as a perturbative extension of the EH action. On the other hand, when approaching the ultraviolet (UV) length scale they become dominant over the Einstein term. To act as truly alternative theory of gravity, the higher order terms have to induce modifications to the solution of the EOM in the non-perturbative regime. This can happen in two scenarios. First, the coefficients are anomalously larger then the cutoff length scale. Second, the length scale is anomalously large so the curvature terms are large. In general, we claim that when\footnote{~$\mathcal{R}^2$ labels all possible combination of quadratic curvature scalars, as in Eq.~(\ref{1.1}).} $ \sum\lambda\mathcal{R}^2\gtrsim R$ or alternatively $\lambda\mathcal{R}\gtrsim1$, significant modifications are induced in the EOM and the theory can be viewed as a modified gravity. We later show explicitly how these relations are the conditions for the EOM to be substantially modified.  

Here, we identify the effective metric - the metric that determines the hyperbolicity of the EOM for propagation of perturbations \cite{yotam}. A more general, related method for identifying the characteristic structure in higher derivative theories is the method of characteristics \cite{Benakli:2015qlh,Izumi:2014loa,Papallo:2017qvl}. This method identify the regions where the Cauchy surface evolves non-uniquely . In this method, the determining criterion for the hyperbolicity of the EOM is the positivity of the principle symbol, which in our case will be strongly related to the demand of having a Lorentzian structure of the effective metric Section (\ref{3.1}). Then, since
 a necessary condition for a hyperbolic EOM is the Lorentzian structure of the effective metric, its non-Lorentzian structure implies a non-hyperbolic EOM. 

Before examining the hyperbolic structure in these extended theories, we address the reader to some simple higher derivative scalar field models. Starting in \cite{Aharonov:1969vu} and later in \cite{yotam, ArmendarizPicon:2005nz, Bruneton:2006gf}. 
 There, it is shown that when the contribution of additional terms in the form $\nabla_\mu\phi\nabla_\mu\phi$ induce significant modification to the EOM, the hyperbolicity effective metric is mo longer guaranteed. 
 
To proceed, we review the "effective metric method", manifested at
\cite{yotam}. The method enables us to identify the effective metric in field space for propagation of perturbations. This method is implemented for higher derivative theories expanded around cosmological Friedman-Robertson-Walker (FRW) backgrounds and for the spherically symmetric Schwarzschild solution. The constraints indicate that the EOM of perturbations are not always hyperbolic when the higher order contributions are significant.

 \section{{Effective metric for quadratic curvature models}}
 
 We will discuss the model in Eq.~(\ref{1.1}) along the same lines of the discussion of the scalar field models. In principle, the difference is that the higher derivative terms $\nabla_\mu\phi\nabla_\nu\phi$, are replaced by their gravitational analogue, namely the $R^2,~ R_{\mu\nu}^2$, $R_{\mu\nu\rho\sigma}^2$ terms. The main complication in comparison to the scalar field models is the index structure and the gauge-redundancy. We overcome these complications by studying gauge-invariant tensor perturbations around FRW spacetime and the Schwarzschild solution.
 
To begin, we consider the action in Eq.~(\ref{1.1}) with the Lagrangian density
\begin{equation}
	\mathcal{L}=\sqrt{{-g}}\left({R}+\lambda_1{R}^2+\lambda_2{R}^{\mu\nu}{R}_{\mu\nu}+\lambda_3{R}^{\mu\nu\rho\sigma}{R_{\mu\nu\rho\sigma}}\right)~.
	\label{L1}
\end{equation}
The vacuum EOM with respect to the background metric are given by,
\begin{equation}
\begin{split}
{G}_{\mu\nu}~&=~{R}_{\mu\nu}-\dfrac{1}{2}g_{\mu\nu}R+(\lambda_2+4\lambda_3)\Box R_{\mu\nu}+\dfrac{1}{2}(\lambda_2+4\lambda_1)g_{\mu\nu}\Box R\\&-~(2\lambda_{1}+\lambda_2+2\lambda_3)\nabla_{\mu}\nabla_{\nu}R+2\lambda_3R_{\alpha\beta\gamma\mu}R^{\alpha\beta\gamma}_{~~~~\nu}+2(\lambda_2+2\lambda_3)R_{\alpha\mu\gamma\nu}R^{\alpha\gamma}\\&-~4\lambda_3R_{\mu\alpha}R_{\nu}^{~\alpha}+ 2\lambda_1RR_{\mu\nu}-\dfrac{1}{2}g_{\mu\nu}(\lambda_1{R}^2+\lambda_2{R}^{\al\beta}{R}_{\al\beta}+\lambda_3{R}^{\al\beta\rho\sigma}R_{\al\beta\rho\sigma}\big)=0~.
\end{split}
\label{GBeom1}
\end{equation}

Eq.~({\ref{GBeom1}}) contain fourth order derivative terms in the metric field like $\nabla_\mu\nabla_\nu R,~\Box R_{\mu\nu}, ~\text{and}~\Box R$. These terms dominate the characteristic structure of the initial value problem and therefore break the hyperbolicity of the EOM \cite{Benakli:2015qlh,Papallo:2017qvl}. In addition, EOM of order higher than two, suffer severe instabilities, such as Ostrogradsky type
instabilities ~\cite{Chen:2012au}. Their Hamiltonian formulation is multivalued in the canonical momentum, which classically resulted in a non-injective time evolution of the initial value problem defined on a Caucy surface \cite{Woodard:2015zca}. We argue that if one wishes to construct an alternative and consistent theory of gravity, the terms containing derivatives that are higher than two in the metric have to be dismissed. As a consequence, the modified EOM would take the differential form as do the Einstein equations take - second order in the metric tensor.

 Different approaches demonstrate how to eliminate the higher than two derivative terms as in Eq.~(\ref{GBeom1}).
One of them, reviewed at \cite{Chen:2012au} suggests the addition of counter terms in the action. These counter terms, when varied with respect to the metric field, cancels out the higher than two derivative terms. Another approach reviewed at \cite{Brustein:2012he} shows that second order field equations can be obtained by redefinition of the tensor modes, in addition to a proper choice of boundary conditions. Then, by specifying the initial-value data on a Cauchy surface the higher than two derivative terms can be eliminated.

Now, in accordance with the above-mentioned and without further details we ignore the fourth order derivative terms. We will be satisfied with the recognition that the possibility of eliminating these terms is exists. The detailed description of this process is out of the scope of this paper. From now and on, our interest would be solely subjected to the examination of the hyperbolicity of the second order field equations.  

To proceed, we expand the metric ${g}_{\mu\nu}$ about the  background $\bar{g}_{\mu\nu}$,
$
{g}_{\mu\nu}=\bar g_{\mu\nu}+ h_{\mu\nu}
$. Then, in accordance, we expand Eq.~(\ref{GBeom1}) to first order in $h_{\mu\nu}$. Again, we emphasize that our aim is to extract the effective metric, we do by isolating the kinetic terms -- terms of the form $\nabla\nabla h$. Then the effective metric $\mathcal{G}_{\al\beta}$ is identified in the form $\mathcal{G}_{\al\beta}\nabla^{\al}\nabla^{\beta} h_{\mu\nu}$.

For our purpose, we evaluate the effective metric for tensor perturbations around maximally symmetric subspace with an Euclidean line element in a 4D spacetime. The generalization for dimensions higher than four is immediate, and would also yield similar results. 

Then, for the isotropic FRW spacetime, tensor perturbations around maximally symmetric spaces are transverse-traceless (TT),
\begin{eqnarray}
h^t_{~i,~}h^i_{~i}, ~\nabla_ih^i_{~j}=0.
\label{Gauge12}
\end{eqnarray}
Where $i=1,2,3$ denote the spatial components. 
 We point out that the choice of tensor modes that are TT with respect to a maximally symmetric subspaces is always possible, as explained in detail at \cite{Higuchi:1986wu}. There it is shown that this choice is independent in the gravitational action, nevertheless it depends on the geometric properties of the spacetime.
 
 This relies on the geometric properties of the space, and not on the gravitational action.  
 
The expansion of Eq.~(\ref{GBeom1}) is made easier by using the following relations.
The first order expansion of the Riemann tensor is given by,
\begin{equation}
\delta{R}^{(1)}_{\mu\nu\rho\sigma}~=~\dfrac{1}{2}\big(\nabla_{\mu}\nabla_{\rho} h_{\nu\sigma}+\nabla_{\nu}\nabla_{\sigma} h_{\mu\rho}-\nabla_{\mu}\nabla_{\sigma} h_{\nu\rho}-\nabla_\nu\nabla_\rho h_{\mu\sigma}\big)~.
\label{Riem12}
\end{equation}
According to the mode choice Eq.~({\ref{Gauge12}}), the expansion of the Ricci tensor to first order is given by,
\begin{equation}
	\delta R_{\mu\rho}^{(1)}~=~\dfrac{1}{2}\left(g_{\nu\sigma}\nabla^\nu \nabla^\sigma h_{\mu\rho}-\nabla^\sigma \nabla_\rho h_{\mu\sigma}\right)~.
	\label{RIC}
\end{equation}
The term $\nabla^\sigma \nabla_\rho h_{\mu\sigma}$, is actually a mass term and has no kinetic contribution to the effective metric. This can be seen by applying the commutation relations of covariant derivatives in terms of the Riemann tensor,	
\begin{gather}
\left[\nabla_\rho, \nabla^\sigma\right]h_{\mu\sigma}=-{R}_{\rho~\mu}^{~\sigma~\al}h_{\alpha\sigma}-{R}^{~\sigma~\alpha}_{\rho~\sigma}h_{\mu\alpha}~,\\
\nabla^\sigma \nabla_\rho h_{\mu\sigma}=-{R}_{\rho~\mu}^{~\sigma~\al}h_{\alpha\sigma}-{R}^{~\sigma~\alpha}_{\rho~\sigma}h_{\mu\alpha}~.
\label{Rcr1}
\end{gather}
 Therefore, in this process we inverted a kinetic term in disguise into a mass terms. Then the kinetic contribution of Eq.~(\ref{RIC}) takes the form
\begin{equation}
\delta R_{\mu\rho}^{(1)}~=~\dfrac{1}{2}g_{\nu\sigma}\nabla^\nu \nabla^\sigma h_{\mu\rho}+\text{mass terms}~.
\label{RIC1}
\end{equation}
The contribution of the Ricci scalar to the effective metric vanishes due to the traceless condition Eq.~({\ref{Gauge12}}). In general, for tensor perturbations Eq.~({\ref{Gauge12}}), the only contribution to the effective metric from Eq.~(\ref{GBeom1}) comes from the Riemann and Ricci tensors.

 Then, one can immediately read the effective metric of the Einstein equations as the coefficient of the kinetic term.
\begin{equation}
\delta\left(2R_{\mu\nu}-g_{\mu\nu}R\right)^{(1)}=g_{\al\beta}\nabla^\al\nabla^\beta h_{\mu\nu}+\text{mass terms}~.
\label{Rv1}
\end{equation}
As expected, $\mathcal{G}_{\al\beta}=g_{\al\beta}$. The effective metric in GR is the background metric with hyperbolic EOM and a well-defined causal structure. 

Then, following the above relations, in addition to the Riemann symmetry properties, we obtain a general expression for the perturbed EOM.
\begin{equation}
\begin{split}
\delta {G}_{ed}~=&~\Box h_{ed}+4\lambda_3R^{abc}_{~~~e}(\nabla_a\nabla_ch_{bd}+\nabla_b\nabla_d h_{ac})~+\\&
2(\lambda_2+2\lambda_3)\Big(R_{adce}\Box h^{ac}+R^{ac}\dfrac{1}{2}\big(\nabla_a\nabla_c h_{de}+\nabla_d\nabla_{e} h_{ac}-\\&\nabla_a\nabla_e h_{dc}-\nabla_d\nabla_c h_{ae}\big)\Big)-
4\lambda_3(R_{da}\Box h_e^a+R_e^a\Box h_{da})+2\lambda_1R\Box h_{de}~.
\label{1411}
\end{split}
\end{equation}

Now, we would like to isolate the kinetic term, so we would have an expression of the form $\mathcal{G}_{\al\beta}\nabla^{\al}\nabla^{\beta} h_{\mu\nu}$. We argue that because of the low symmetry demonstrated by such models, an implicit expression of the effective metric in terms of the curvature tensors is very complicated and is also irrelevant for our purpose. In contrast to other theories, such as Lovelock theories, where it is shown \cite{yotam} that due to the high degree of symmetry of the fully anti-symmetric Lovelock Lagrangians, an implicit expression of the effective metric in terms of the curvature tensors takes a simple form and, its derivation is significantly simpler. Here, we will be satisfied with an explicit calculation of the curvature tensors about a specific background solutions. The calculation is carried out by considering the non-vanishing components of the Riemann tensor in field space as we explain in detail below.

\subsection{The effective metric in FRW spacetime}\label{3.1}
The homogeneous and isotropic 4D FRW spacetime  with the line element
\begin{equation}
ds^2~=~-dt^2+a(t)^2\gamma_{{i}{j}}dx^{{i}}dx^{{j}},
\label{frw}
\end{equation}
where $i,j=1,2,..,D-1$ denote spatial components, and $\gamma_{{i}{j}}dx^{{i}}dx^{{j}}$ is the Euclidean line element.
This type of spacetime solves action Eq.~(\ref{1.1}) in the presence of matter. A detailed description of the solutions is irrelevant for our purpose, and can be found at \cite{Clifton:2011jh,Carroll:2004de,Nojiri:2010wj}. Here, we will need to be  with the existence of solutions describing a universe undergoing a decelerated or accelerated expansion.

As previously mentioned, we evaluate the effective metric here for tensor perturbations, which in the case of the FRW backgrounds are characterized by
	\begin{equation}
	h^{\bar{a}}_{~\bar{a}}=0~,~~~~~~~~~
	\qquad
	h^{\bar{a}}_{~t}=0~,~~~~~~~~~
	\qquad
	\nabla_{\bar{a}}h^{\bar{a}}_{~\bar{b}}=0~.
	\label{13}
	\end{equation}
	where barred indices $~\bar{a}, \bar{b}=1,2,3$,  denote spatial components. The non-vanishing components of the Riemann tensor are the following,
	\begin{equation}
\mathcal{R}^{\bar{a}\bar{b}}_{\bar{c}\bar{d}}~=~ H^2\delta^{\bar{a}\bar{b}}_{\bar{c}\bar{d}}~,~~~~~~~~~~~~~~~~~
\qquad
\mathcal{R}^{t\bar{a}}_{t\bar{c}}~=~\delta^{\bar{a}}_{\bar{c}}\frac{\ddot{a}}{a}~.
\label{41}
\end{equation}	
	Where $(\dot{a}/a)^2=H^2$ and ${\ddot{a}}/{a}=H^2+\dot H$. Now, by taking the Riemann tensor in terms of the background metric we are able to derive an explicit expression of the effective metric.

We now wish to explain how to evaluate the effective metric for tensor perturbations about the FRW background. For simplicity, we first consider the term  $R^{ab}_{~~ce}(\nabla_a\nabla^ch_{bd}+\nabla_b\nabla_d h_{a}^c)$	
~(taken from Eq.~(\ref{1411})). The evaluation is carried out in 4 steps.   

\textbf{1.} Express the Riemann tensor in terms of its different background values {metric Eq.~(\ref{41}).}
\begin{equation}
	\begin{split}
&R^{ab}_{~~ce}(\nabla_a\nabla^ch_{bd}+\nabla_b\nabla_d h_{a}^c)~=\\
&{R}^{t\bar{b}}_{~~t\bar{e}}(\nabla_t\nabla^th_{\bar{b}\bar{d}}+\nabla_{\bar{b}}\nabla_{\bar{d}} h_{t}^t)+{R}^{\bar{a}\bar{b}}_{~~\bar{c}\bar{e}}(\nabla_{\bar{a}}\nabla^{\bar{c}}h_{\bar{b}\bar{d}}+\nabla_{\bar{b}}\nabla_{\bar{d}} h_{\bar{a}}^{\bar{c}})~=\\&
\delta^{\bar{b}}_{\bar{e}}\frac{\ddot{a}}{a}\nabla_{t}\nabla^th_{\bar{b}\bar{d}}+H^2\delta^{\bar{a}\bar{b}}_{\bar{c}\bar{e}}\left(\nabla_{\bar{a}}\nabla^{\bar{c}}h_{\bar{b}\bar{d}}+\nabla_{\bar{b}}\nabla_{\bar{d}} h_{\bar{a}}^{\bar{c}}\right)~.
	\end{split}
	\label{18}
\end{equation}

	\textbf{2.} Perform contraction and apply mode choice Eq.~(\ref{13}).
	\begin{equation}
	\begin{split}
&\frac{\ddot{a}}{a}\nabla_{t}\nabla^th_{\bar{e}\bar{d}}+H^2\left(\nabla_{\bar{a}}\nabla^{\bar{a}}h_{\bar{e}\bar{d}}+\nabla_{\bar{e}}\nabla^{\bar{b}}h_{\bar{b}\bar{d}}+\nabla_{\bar{e}}\nabla^{\bar{d}}h_{\bar{c}}^{\bar{c}}+\nabla_{\bar{c}}\nabla^{\bar{b}}h_{\bar{e}}^{\bar{c}}\right)=\\&
\frac{\ddot{a}}{a}\nabla_{t}\nabla^th_{\bar{e}\bar{d}}+H^2\left(\nabla_{\bar{a}}\nabla^{\bar{a}}h_{\bar{e}\bar{d}}+\nabla_{\bar{c}}\nabla^{\bar{b}}h_{\bar{e}}^{\bar{c}}\right)~.
\end{split}
	\label{181}
\end{equation}

\textbf{3.} Dismiss mass terms by using the commutation relations of covariant derivatives Eq.~(\ref{Rcr1}). 
	\begin{equation}
\frac{\ddot{a}}{a}\nabla_{t}\nabla^th_{\bar{e}\bar{d}}+H^2\nabla_{\bar{a}}\nabla^{\bar{a}}h_{\bar{e}\bar{d}}+\text{mass terms}~.
	\label{182}
	\end{equation}
	
	\textbf{4.} Gather the different time-time and space-space derivative terms separately, then move into covariant representation.
	\begin{equation}
\mathcal{G}^{\alpha\beta}\nabla_\al \nabla_{\beta}h_{\bar{e}\bar{d}}~=~\frac{\ddot{a}}{a}g^{tt}\nabla_{t}\nabla_th_{\bar{e}\bar{d}}+H^2g^{\bar{a}\bar{c}}\nabla_{\bar{a}}\nabla_{\bar{c}}h_{\bar{e}\bar{d}}~.
	\label{183}
	\end{equation}
In this specific example there was only one contribution for each derivative term, so step {4} was trivial. This will not be the case when more terms are included as in Eq.~(\ref{1411}).

Following the instructions above, we obtain the effective metric about FRW spacetime for EOM Eq.~(\ref{GBeom1}).

 The time-time component
	\begin{equation}\begin{split}
	&\mathcal{G}^{tt}\nabla_t\nabla_th_{\bar{e}\bar{d}}~=\\&
	g^{tt}\left(1+\dot{H}(2\lambda_3+3\lambda_2+12\lambda_1)+H^2(-10\lambda_3+5\lambda_2+24\lambda_1)\right)\nabla_t\nabla_th_{\bar{e}\bar{d}}~.
	\end{split}
	\label{ttm}
	\end{equation}	
	The space-space component.
	\begin{gather}\begin{split}
	&	\mathcal{G}^{\bar{a}\bar{c}}\nabla_{\bar{a}}\nabla_{\bar{c}}h_{\bar{e}\bar{d}}~=\\&
	g^{\bar{a}\bar{c}}\left(1+\dot{H}(-6\lambda_3+\lambda_2+12\lambda_1)+H^2(-10\lambda_3+5\lambda_2+24\lambda_1)\right)\nabla_{\bar{a}}\nabla_{\bar{c}}h_{\bar{e}\bar{d}}~.
	\label{ssm}
	\end{split}
	\end{gather}
	First, it is clear the effective metric has no dependence on the different graviton polarization components $h_{\bar{e}\bar{d}}$, the reason stems from the fact that the tensor modes we are looking at Eq.~(\ref{13}) are defined about the spatial maximally symmetric subspace of the FRW spacetime. In contrast, if the tensor modes were defined about different maximally symmetric subspaces, the effective metric was polarization dependent. For example, if we are looking at a specific graviton $h_{ed}$ with polarization components $(e,d)$, a term like $R^{abc}_{~~~e}$ in the effective metric Eq.~(\ref{1411}) can have a polarization dependent contribution, since it takes different values for a different polarization components. 
	 
	It is important to notice that the effective metric time and the space components have a different proportionality factors. So, in total, the effective metric is not proportional to the background metric $g^{\mu\nu}$, thus, enabling the breakdown of the Lorentzian structure.
In particularly, since the multiplying factors in parenthesis of the respective time Eq.~(\ref{ttm}) and space Eq.~(\ref{ssm}) components can have a different sign, then in general the effective metric will not be Lorentzian.
 Alternatively, if these factors are of the same sign, then the effective  metric will Lorentzian.

  In addition, we note that the factors multiplying $H^2$ are identical in both the spatial and temporal components. We also note that in contrast to $\dot{H}$, the factors multiplying $H^2$ are identical in both the spatial and temporal components. Thus, the only deviation from the standard Lorentzian pattern can be attributed to the magnitude of $\dot{H}$.  

We now turn to discuss the results and to examine the conditions under which the EOM are non-hyperbolic. We do that by considering the different combinations of the higher order terms in action Eq.~(\ref{1.1}). We show that the conditions for the well-posedness of the perturbed EOM are in general, depends on the specific values of the metric parameters, $a(t)$ in FRW spacetime and the mass the Schwarzschild spacetime. The classification of the different combinations will allow us to determine which theories yield an ill-posed initial value problem and therefore are inapplicable as an alternative theory of gravity.

It is important to mention that by using the Gauss-Bonnet (GB) invariant, the Riemann squared term in action Eq.~(\ref{1.1}), can be solely expressed in terms of the Ricci tensor and the scalar curvature. This is due to the fact that the GB term, namely $ \mathcal{L}_{GB}=R^2-4R_{\mu\nu}R^{\mu\nu}+R_{\mu\nu\rho\sigma}R^{\mu\nu\rho\sigma} $  is a total derivative in 4D, and therefore has no influence on the solutions of the EOM. However, though the EOM with respect to the first variation of action Eq.~(\ref{1.1}) yields identical solutions, the EOM for perturbation with respect to the second variation are essentially different.  
This can easily be seen in Eqs.~(\ref{1411}),(\ref{18}),(\ref{181}), where the contribution of the Riemann tensor to the perturbed EOM has an additional terms, involving the contribution from the different degrees of freedom.  In general, these extra terms would not be present if one applies the GB identity. Therefore, we will consider action Eq.~(\ref{1.1}) in its explicit form. 

We recall again that our purpose is to show that when the higher derivative terms are large, the hyperbolicity effective metric is mo longer guaranteed and therefore, the extended theory cannot
be viewed as a consistent theory of gravity.

\subsubsection{R$^2$ theories}
We consider the Lagrangian term of the form $\mathcal{L}=R+\lambda_1R^2$.
In this form, the effective metric is immediately found by taking $\lambda_{1}\neq 0,~\lambda_2=\lambda_3=0$ in Eqs.~(\ref{ttm}),(\ref{ssm}). Thus obtaining,
\begin{gather}
\mathcal{G}^{tt}=g^{tt}\left(1+\lambda_1\left(12\dot{H}+24H^2\right)\right)~,\\
\mathcal{G}^{\bar{a}\bar{c}}=g^{\bar{a}\bar{c}}\left(1+\lambda_1\left(12\dot{H}+24H^2\right)\right)~.
\end{gather}
It is clear that since the factors multiplying $g^{tt}$ and $g^{\bar{a}\bar{c}}$ are identical, the effective metric is proportional to the background metric and therefore its signature is always Lorentzian, no matter the magnitude of $\lambda_1$. 

This result is not surprising at all, since as already demonstrated by Starobinsky at \cite{Starobinsky:1980te} 
 (for a detailed review \cite{DeFelice:2010aj}), the extensions of the EH action by terms that are powers of Rici scalar is equivalent to the insertion of scalar potential in the action or alternatively as the addition of mass terms, which in both cases has no influence on the hyperbolicity of the EOM for tensor perturbations as shown at \cite{Alvarez-Gaume:2015rwa}. We therefore conclude that extensions of the form $R^n$ has no influence on the dynamics of the perturbed EOM. 

\subsubsection{$R_{\mu\nu\rho\sigma}R^{\mu\nu\rho\sigma}$ extensions}
The Lagrangian term we are considering is 
$\mathcal{L}=R+\lambda_3R_{\mu\nu\rho\sigma}R^{\mu\nu\rho\sigma}$.
Replacing $\lambda_{3}\neq 0,~\lambda_1=\lambda_2=0$ in Eqs.~(\ref{ttm}),(\ref{ssm}).
The corresponding effective metric is,
\begin{gather}
\mathcal{G}^{tt}=	g^{tt}\left(1+\lambda_3H^2\left(2\dfrac{\dot{H}}{H^2}-10\right)\right)~,\\
\mathcal{G}^{\bar{a}\bar{c}}=g^{\bar{a}\bar{c}}\left(1-\lambda_3H^2\left(6\dfrac{\dot{H}}{H^2}+10\right)\right)~.
\label{ricci12}
\end{gather}
First, it is clear that for $\lambda_3=0$ we recover the Einstein result, so the effective metric becomes the background metric. For $\lambda_3\neq0$ the factors multiplying $g^{tt}$ and $g^{\bar{a}\bar{c}}$ are different and in some cases can have opposite signs.
In accordance with the assumption mentioned at Section (\ref{2}), we assume that the correction terms are non-negligible; $\lambda_iH^2\gtrsim1$.
Then, the hyperbolicity is governed by the $\dot{H}/H^2$ term. So, for $\lambda_{3}>0$~\footnote{~The case of $\lambda_i<0$ is identical to the $\lambda_i>0$ case.} the signature becomes non-Lorentzian in the range ${\dot{H}}/{H^2}>5 ~,~{\dot{H}}/{H^2}<-5/3$.
  
  To understand the significance of the conditions, let us consider a solution of decelerated expansion \cite{Carroll:2004de,Nojiri:2010wj} of the form $a(t)\sim t^\alpha$ with $0<\alpha<1$. In this case, $\dot H = - H^2/\alpha $. Then, the signature becomes non-Lorentzian for $\alpha$ parameters $\al<{3}/{5}$.  
  This condition (when $\lambda_3 H^2\gtrsim 1$) rules out a large range of the parameter space of decelerated expanding isotropic solutions.

 Another way of interpreting these results is by examining the influence of these higher derivative terms on the propagation speed of perturbations. For simplicity, we consider a radially propagating perturbation. The EOM are given by
\begin{equation}
	\mathcal{G}^{ab}\nabla_a\nabla_b h_{\mu\nu}~=~	\mathcal{G}^{tt}\nabla_t^2{h}_{\mu\nu}+	\mathcal{G}^{rr}\nabla_r^2{h}_{\mu\nu}~.
\end{equation}
\begin{figure}[t]
	\centering
	\includegraphics[width=1.15\linewidth, height=0.323\textheight]{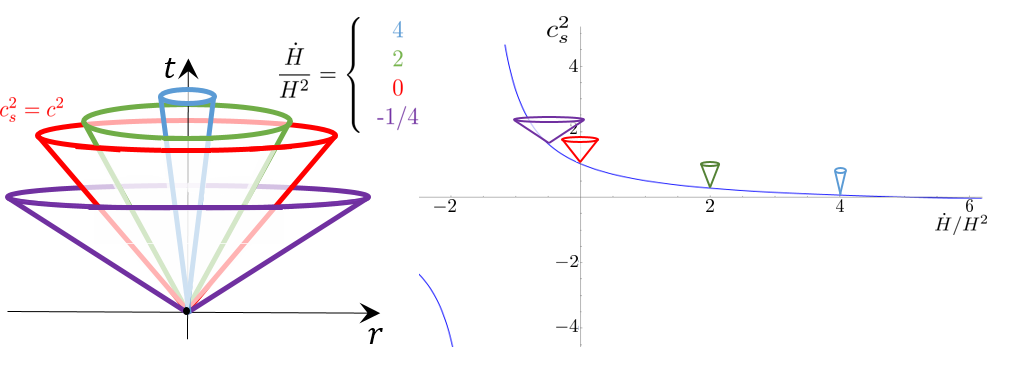}
	\caption[p]{ {\footnotesize The effective speed of gravitational\vspace{-.1cm} perturbations as a function of $\dot{H}/H^2$. The hyperbolic regions where $-5/3<\dot{H}/H^2<5$ with $c_s^2>0$\vspace{-.1cm} are drown, (the specific values were chosen for $\lambda H^2=1$). The effective light cone "shrinks"\vspace{-.1cm}with increasing $\dot{H}/H^2$ until it becomes fully degenerated at $\dot{H}/H^2=5,-5/3$ making $c_s$ \vspace{-.1cm}imaginary.}}
	\label{fig:cc3}
\end{figure} 
The null geodesics define the effective speed of propagation $ c_s^2=-{\mathcal{G}_{tt}}/{\mathcal{G}_{rr}} $. Then $c_s^2$ is been substantially modified when the higher derivative terms are significant. An interesting observation is that one can still have a well-defined causal structure, even though the propagation speed exceeds the speed of light Figure (\ref{fig:cc3}). Particularly interesting are the conditions where $c_s$ becomes imaginary, that are identical to the non-Lorentzian conditions mentioned above. The imaginary regime of $c_s$ indicates on a causal structure that is badly defined and on the degeneracy of the Cauchy surface.

We now wish to explain the condition $\lambda H^2\gtrsim1$. Recalling that  significant modifications to the EH action are induced when the higher order terms satisfies $\lambda\mathcal{R}\gtrsim1$  Sec.(\ref{2}). Then, since in FRW spacetime $\mathcal{R}\sim H^2$, the condition, $\lambda H^2\gtrsim1$ coincides with the above relations. The interpretation in terms of the $\al$ parameter implying that for  for $a(t)\sim t^{\al}$ we have $\lambda \frac{\al^2}{t^2}\gtrsim 1$ which always valid at some early times. Alternatively, at late times, one can require the coefficient $\lambda$ to be large such that  $\lambda H^2\gtrsim1$ is valid.

In general, we claim that the breakdown of the causal structure can be avoided if one constraints the correction terms, such that they are subdominant. For example, by demanding $\lambda H^2<1$, or alternatively $\frac{\dot{H}}{H^2}\ll1$. However, this implies that cutoff scale of the theory is set by the correction terms and not by the Planck scale or other independent high scale. 
 
\subsubsection{$R_{\mu\nu}R^{\mu\nu}$ theories}
For the contributions of the form $\lambda_2R_{\mu\nu}R^{\mu\nu}$, we obtain the effective metric by taking $\lambda_1=\lambda_3=0$ in Eqs.~(\ref{ttm}),(\ref{ssm}).
\begin{gather}
\mathcal{G}^{tt}=	g^{tt}\left(1+\lambda_2H^2\left(3\dfrac{\dot{H}}{H^2}+5\right)\right)\label{ricci34}~,\\
\mathcal{G}^{\bar{a}\bar{c}}=g^{\bar{a}\bar{c}}\left(1+\lambda_2H^2\left(\dfrac{\dot{H}}{H^2}+5\right)\right)~.
\label{ricci3}
\end{gather}
In similar to the previous example. When the multiplying factors in parenthesis of Eqs.~(\ref{ricci34}),(\ref{ricci3}) have a different sign the signature is non-Lorentzian. We assume again that $\lambda_2H^2\gtrsim1$, then the hyperbolicity of the EOM is governed by the $\dot{H}/H^2$ term.
Therefore, when $\lambda_2>0$ the signature becomes non-Lorentzian when $-5<{\dot{H}}/{H^2}<-5/3$.

We interpret these results again by considering a decelerated expansion solution with $a(t)\sim t^\alpha$ and $0<\alpha<1$. In this case, $\dot H = - H^2/\alpha $. Then, the signature becomes non-Lorentzian for $\alpha$ parameters $1/5<\al<3/5$.  
This condition (when $\lambda_2 H^2\gtrsim 1$) rules out the radiation dominated isotropic solutions.

\subsubsection{$ R_{\mu\nu}R^{\mu\nu}+R_{\mu\nu\rho\sigma}R^{\mu\nu\rho\sigma} $ extensions}
We consider the case where ${\lambda_2=\lambda_3\text{~and~} ~\lambda_1=0}$, then the Lagrangian term becomes $ \mathcal{L}~=~R+\lambda\left(R_{\mu\nu}R^{\mu\nu}+R_{\mu\nu\rho\sigma}R^{\mu\nu\rho\sigma}\right)  $.
The corresponding effective metric is 
\begin{gather}
\mathcal{G}^{tt}~=~g^{tt}\left(1+5\lambda H^2\left(\dfrac{\dot{H}}{H^2}-1\right)\right)\\
\mathcal{G}^{\bar{a}\bar{c}}~=~g^{\bar{a}\bar{c}}\left(1-5\lambda H^2\left(\dfrac{\dot{H}}{H^2}+1\right)\right)
\end{gather}
Again, under the same assumptions ($\lambda H^2\gtrsim1$). We conclude that the signature is non-Lorentzian, thus making the EOM non-hyperbolic when $\dot{H}/H^2>-1$. The interpretation of this result in terms of scale factor in the form $a(t)\sim t^{\al}$ with $\al>0$ shows that for all inflation solutions with $\al>1$ parameters the EOM for tensor perturbation are non-hyperbolic.   

 \subsection{ $C_{\mu\nu\rho\sigma}C^{\mu\nu\rho\sigma}$ theories}
 We now consider separately the contribution of the quadratic Weyl tensor to the EH action, this terms are considered at \cite{Donoghue:2018izj,Lu:2015cqa}
 \begin{equation}
 	\mathcal{S}=\dfrac{1}{16\pi l_p^2}\int d^4x\sqrt{-g}\big({R}+\lambda C_{\mu\nu\rho\sigma}C^{\mu\nu\rho\sigma}\big)~.
 	\label{weyl}
 \end{equation}
 The Weyl tensor is given by
 \begin{equation}
 \begin{split}
 	C_{\mu\nu\rho\sigma}~=&~R_{\mu\nu\rho\sigma}-\dfrac{1}{2}\left(R_{\mu\rho}g_{\nu\sigma}+R_{\nu\sigma}g_{\mu\rho}-R_{\mu\sigma}g_{\nu\rho}-R_{\nu\rho}g_{\mu\sigma} \right)+\\&~\dfrac{1}{6}\left(g_{\mu\rho}g_{\nu\sigma}-g_{\mu\sigma}g_{\nu\rho}\right)R~,
 	 \end{split}
 \end{equation}
 and the quadratic Weyl tensor is
\begin{equation}
C_{\mu\nu\rho\sigma}C^{\mu\nu\rho\sigma}~=~{R}^{\mu\nu\rho\sigma}{R_{\mu\nu\rho\sigma}}-2{R}^{\mu\nu}{R_{\mu\nu}}-\dfrac{1}{6}R^2~.\label{35}
\end{equation} 

 The effective metric is easily obtained from Eqs.~(\ref{ttm}),(\ref{ssm}) by replacing $\lambda_3=\lambda,~\lambda_2=-2\lambda, ~\lambda_1=-\frac{1}{6}\lambda$, thus
\begin{gather}
\mathcal{G}^{tt}~=~
g^{tt}~\left(1-\lambda H^2\left(6\dfrac{\dot{H}}{H^2}+24\right)\right)
\\	\mathcal{G}^{\bar{a}\bar{c}}~=~\left(1-\lambda H^2\left(10\dfrac{\dot{H}}{H^2}+24\right)\right)
\label{ssm1}
\end{gather}
The, for $\lambda H^2\gtrsim1$ We conclude that the signature is non-Lorentzian when $-4<\dot{H}/H^2<-2.4$. The interpretation of this result in terms of scale factor in the form $a(t)\sim t^{\al}$ with $\al>0$ shows that when $\frac{1}{4}<\al<\frac{5}{12}$ the EOM for perturbations are non-hyperbolic. Which again, rules out partially the range describing decelerated expansion solutions. 
 
\subsection{Effective metric in Schwarzschild spacetime }\label{s3.2}
Here, we apply the effective metric method for the static spherically symmetric Schwarzschild BH solution\cite{Lu:2015cqa,Stelle:1977ry}. Particularly interesting is the behavior of the EOM for perturbation in large curvature background, and especially near the horizon.

The static spherically symmetric Schwarzschild spacetime is a solution 
of the modified action Eq.~(\ref{1.1}). This observation is immediately seen by exploiting the properties of the GB invariant Eq.~(\ref{35}). Indicating that the Lagrangian Eq.~(\ref{1.1}) can be decomposed into two independent contributions. In particularly by $\al R^2+\beta R_{\mu\nu}R^{\mu\nu}~$
\cite{Stelle:1977ry}. As a consequence, any solution that satisfies the vacuum Einstein equation will also satisfies the modified EOM.

To proceed, the static spherically symmetric Schwarzschild BH solutions are
\begin{equation}
ds^2=-\left(1-\dfrac{2M}{r}\right)dt^2+\left(1-\dfrac{2M}{r}\right)^{-1}dr^2+r^2d\Omega^2,
\label{BHmet}
\end{equation}
where $d\Omega^2$ is the 2D unit sphere. 
The non-vanishing components of the Riemann components are the following,
\begin{equation}
\mathcal{R}^{tr}_{tr} =-\dfrac{f''(r)}{2}~,
~~~~~
\mathcal{R}^{ij}_{kl}=-\dfrac{f(r)}{r^2}\delta^{ij}_{kl}~,
~~~~~
\mathcal{R}^{~\alpha i}_{\alpha j}=-\dfrac{f'(r)}{2r}\delta^i_j~.
\label{f3}
\end{equation}
Where $f(r)=(1-2M/r)$, the indices  $i, j, k, l=1,2$ denote angular coordinates and $\alpha = t, r$. 
We are interested in the tensor modes about the maximally symmetric subspace. Recalling that such choice is always possible when expanding about the Euclidean line element Eq.~(\ref{BHmet}). They are given by
\begin{equation}
\begin{split}
&h^{\alpha\beta}  =0~,
~~~~
h^{\alpha i}  =0~,\\
~~~
&h^{i}_{~i}  =0~,
~~~~~
\nabla_i h^{ij} =0~.
\end{split}
\label{g24}
\end{equation}

The calculation of the effective metric is carried out in a similar way to the one performed in the previous FRW examples. The main difference compared to the FRW case is that the Ricci tensor and the scalar curvature are vanishing, enabling us to perform the calculation with more ease. the effective metric component are give by
\begin{gather}
\mathcal{G}_{\al\al}= g_{\alpha\alpha}\left(1-\lambda\left(2\dfrac{f'(r)}{r}-6\dfrac{f(r)}{r^2}\right)\right)~,		
\label{BH5}\\
\mathcal{G}_{kk}= g_{kk}\left(1-2\lambda\dfrac{f(r)}{r^2}\right)~.
\label{BHef5}
\end{gather}
As in the FRW case, symmetry results in an equal effective metric for all polarizations. First, it is clear that the multiplying factors in parenthesis are different, implying the possibility for a deviation from the standard Lorentzian pattern. Second, since both $f(r),f'(r)>0$, the issue is whether the factor multiplying $\lambda$ in Eq.~(\ref{BH5}) can change it sign. We derive an explicit conditions for the non-Lorentzian signature by substituting the metric function $f(r)$ in the above equations. The corresponding effective metric 
\begin{gather}
\mathcal{G}_{\al\al}= g_{\alpha\alpha}\left(1-\dfrac{\lambda M}{r^3}\left(6\al-16\right)\right)~,		
\label{BH56}\\
\mathcal{G}_{kk}= g_{kk}\left(1-\dfrac{\lambda M}{r^3}\left(2\al-1\right)\right)~.
\label{BHef56}
\end{gather}
Where $\al=\frac{r}{M}$ and $\lambda>0$ \footnote{~The case of $\lambda<0$ is identical to the $\lambda>0$ case.}. The unitless coefficient $\frac{\lambda M}{r^3}$ was extracted, since when it satisfies  $\frac{\lambda M}{r^3}\gtrsim 1$  the higher derivative can induce significant modifications to the EH action. Then, noting that $\mathcal{R}\sim \frac{M}{r^3}$ \footnote{In general, we consider the largest contribution from $\mathcal{R}$, originates in $R_{abcd}R^{abcd}=\frac{48M^2}{r^6}$.}, the above relations coincides with  $\lambda\mathcal{R}\gtrsim 1$.

To proceed, it is clear is that for any value of $\lambda$ and at large enough distances the effective metric reduces to the background Schwarzschild metric, so it is Lorentzian in this region of large $r$. Therefore, the effective metric can be non-Lorentzian only for a smaller values of $r$, where the corrections due to the higher derivative terms are large. Here we find that the initial value problem is ill-posed when $2<\al<2.75$. The conclusion is that there exist a physical region exterior to the BH horizon $2M<r<2.75M$  where perturbation are not propagate in a causal way.


\section{Summary and Conclusions}

In this paper we examine whether modifications to the EH action in the form of quadratic curvature terms can provide an alternative theory of gravity.
We analyzed the causal structure of the EOM for perturbation about the FRW and Schwarzschild spacetimes, and found it to be violated when the higher derivative terms are comparable or larger than the Einstein term, namely $\lambda\mathcal{R}\gtrsim1$.
Therefore, we conclude that these theories cannot constitute a modified gravity theory. On the other hand, when these terms are small and therefore act as a perturbative correction of the Einstein term, the EOM are hyperbolic because they are hyperbolic for GR.

First, we review of the effective metric method. Then, following the formalism \cite{yotam}, we derive the effective metric for the gravitational case when quadratic terms are included in the EH action. We emphasize that our interest was mainly subjected to EOM that are in the form of Einstein equations - second order in the background field.

Then, we performed explicit calculations of the effective metric about the FRW cosmological spacetime and the spherically symmetric Schwarzschild solution, when different combination of extensions are included in the EH action Sec.~(\ref{3.1}). We found that in general, the Lorentzian pattern of the effective metric in FRW spacetime Eqs.~(\ref{ttm})(\ref{ssm}) is governed by the magnitude of $\dot{H}$. Then, we analyzed the results by assuming  $\lambda H^2\gtrsim1$, then the non-Lorentzian conditions were expressed in terms of the ratio $\dot{H}/H^2$. The results were also interpreted in terms of the specific values of $\al$ parameters.    
We find that due to the presence of the higher order terms the EOM for perturbations are no longer hyperbolic. Indicating on the violation of causality for a large range of $\al$ parameters describing a universe undergoing an inflation or deceleration.

We remark again that the hyperbolicity of the EOM can be ensured when the cutoff scale of the theory is set by the correction terms, thus making them subdominant. For example, if one impose a cutoff on the theory such that  $\lambda H^2<1$, or alternatively if one constraints $\dot{H}$ (allowing $\lambda H^2$ to be large), then the correction terms become subdominant and can be considered as a perturbative extension of GR. Thus, making the EOM for perturbations hyperbolic. 

Finally, we demonstrated the method for the Schwarzschild solution. The analysis was carried out along the same lines of the FRW spacetime. We found that when the correction term are large $\frac{\lambda M}{r^3}\gtrsim1$, then, outside the horizon, the effective metric becomes non-Lorentzian for  $2M<r<2.75M$. The conclusion, in similar to the FRW case, indicating the possibility of causality violation when the higher derivative terms are treated in the non-perturbative regime. 


To finish, the conclusion is that the breakdown of predictability in higher derivative theories indicates that these theories can only be considered as a perturbative extension of the EH action, and not as a true alternative to GR. 



\section*{Acknowledgments}
I would like to Ramy Brustein for the useful comments and the valuable discussions. Yoav Zigdon and Dror Sherf for the comments. 
 The research was supported by the Negev Hi-Tech scholarship, and by the Israel Science Foundation grant no.
 1294/16.

\end{document}